\documentstyle [epsfig,aps,preprint] {revtex}

\begin{document}

\title{Optical characterization of chevron texture formation in nematic 
electroconvection}

\author{ H. Amm, R. Stannarius, and A. G. Rossberg\dag}

\address{ Universit\"at Leipzig, Fak. Physik und Geowiss., Linn\'{e}str. 5,
          D-04103 Leipzig, Germany\\
          \dag Universit\"at Bayreuth, Physikalisches Inst.
          D-95440 Bayreuth, Germany}

\maketitle

\centerline{(Received \today)}

\begin{abstract}

  We characterize the director structures in electroconvection 
  patterns of nematic liquid crystals by means of polarising microscopy. 
  This study was stimulated by a theory put forward by
  Rossberg et al.\cite{Rossberg} who propose a mechanism
  for chevron texture formation. 
  We characterize the relation between the wave vector modulation and
  in-plane twist modulations of the director field.
  In addition to the standard optical setup, a circular analyzer is used.
  The results provide new insight into the interplay of director and
  wave vector field 
  leading to the chevron texture and fluctuating conductive rolls.
  Our experimental results confirm the theoretical predictions.

\end{abstract}
\pacs{PACS numbers:
  47.20-k, 
  61.30-v  
  }

\section*{INTRODUCTION}

Electroconvection (EC) in nematics is a well known
standard system for dissipative pattern formation.
It provides a rich scenario of dissipative structures
which can be readily controlled by experimental parameters as electric
field strength and frequency.
The study of EC in nematics has attracted much interest during the
last decades. The hydrodynamic and electric equations
involved are well known. Observation of the textures is straightforward 
by means of optical microscopy. 
Much progress has been made during the last
decades in the theoretical understanding of the pattern selection process.
The basic classification and physical mechanism of the different patterns 
are well understood.

The system has still retained its attractivity 
for scientific research \cite{Kramer}. Because of its high complexity, 
many features have found theoretical explanation only recently or still 
lack sufficient understanding.
While the periodicity and orientation of the optical patterns can be
easily studied with transmission microscopy, the description of the
director field in the convection structures is still not fully revealed.
Most experiments with EC so far have been restricted to observations
in unpolarized light or in a polarizing microscope with polarizers along
the director easy axis $\vec n_0$. 
In that geometry, periodical tilt modulations
of the director generate a lense effect and produce a periodically
modulated transmission texture which maps the director pattern.
The optical characteristics can be calculated numerically using 
Fermat's principle, and a relation
between the observed texture and the director tilt profile
can be established at least for the simplest convection roll patterns.

One of the more complex patterns that can be observed is the ''chevron
texture'' which is named after a characteristic periodic modulation
of the roll orientation as it can be seen in Figs.~\ref{image2},\ref{image3}. 
An explanation of the formation of such chevron patterns has
been suggested recently \cite{Rossberg}.  In order to describe the
appearance of chevron patterns from a homogeneously defect-chaotic
state, the authors propose a set of model equations of the
reaction-diffusion type, which is well known to lead to spatially
modulated patterns ({\em Turing}-patterns) under appropriate conditions. 
The roles of fast and the slowly diffusing species in these model equations 
are played by the in-plane director reorientation (in the present case
a twist which is superimposed to the usual periodic tilt structure of
the convection rolls) and the orientation of the wave vector of the
rolls, respectively.  While the ''diffusion'' of the director twist is
mediated by the bend elastic forces of the liquid crystal, the wave vector
diffuses by the irregular motion of individual defects, which is
characteristic for the defect-chaotic state. The theoretical description
has been developed originally for systems with weakly broken isotropic
symmetry, as represented by EC in homeotropic cells under the influence
of a lateral magnetic field. In that case, the proposed mechanism leads
to a periodic bend deformation of the director accompanied by a
corresponding modulation of the wave vector of the roll patterns.
In the system studied here, EC in planar cells, the situation is more
complex due to the preferred alignment axis of the director at the glass
plates. The director remains fixed at the upper and lower cell boundaries
while it twists out of the alignment axis in the cell middle. A twist
deformation along the cell normal is expected to arise.

The local director and the wave vector of the rolls interact by a
systematic drift of defects which is induced by the director twist and
leads to a change of the wave vector. In addition, director and wave
vector are coupled by a hydrodynamic torque on the director, 
which must generally be expected when the director is not
aligned perfectly normal to the rolls. For liquid crystal
convection this torque seems generally to increase
misalignment \cite{Lindner,RossbergII}. If the director
would play a rather passive roll in chevron formation, one might expect
that director and wave vector are deflected in opposite directions.
In contrast, according to the proposed model, the linear mode associated
with the chevron formation must contain both the well known
long-range modulation of the wave vector and a modulated director
twist in phase with it. In fact the amplitudes of the two
components are expected to be of similar size.

Within this study, we have focused our investigation on the detection of
such out-of-plane (out of the tilt plane between surface alignment and
cell normal) twist director deflections,
in addition to the standard characterization of the dielectric convection
structures. With modified optical techniques,
we analyse the complex structure of the director field and its relation to 
the wave vector field of the convection rolls.
In standard planar cells with $\approx$ 20 to 100$\mu m$
cell gap, small director twist cannot be detected in the conventional
experiment. The polarization plane of incident light is guided
adiabatically with the director twist along the cell normal, and it is
rotated back before leaving the nematic layer. This phenomenon known as
wave guiding is effective as long as the so called {\em Mauguin} limit
holds \cite{mauguin},
that is the optical axis of the anisotropic medium changes its 
orientation in space sufficiently slowly on the wavelength scale.
Therefore, the optical transmission intensity is practically
uninfluenced by moderate director twist
\cite{cladis,gerber}. In conventional experiments,
only the tilt profile of the EC patterns is accessed,
information on the director field is incomplete.
Possible director twist is usually neglected in the description of EC 
patterns.
The actually three-dimensional director field has not been characterized 
yet because of the experimental limitations.

However, it is well known that a small modification of the optical setup,
the insertion of a quarterwave plate at $45^\circ$ orientation to the
polarizers raises the degeneracy between clockwise and counterclockwise 
twisted domains. It can be used to visualize twist deformations in the 
nematic sample \cite{grigutsch}. 
Using such a phase plate, we record the contrast modulation in the
cell plane which is a measure of the director twist pattern. We compare
these twist modulations with the arrangement of the convection rolls.

\section*{SUBSTANCES AND EXPERIMENTAL SETUP}

We use glass cells with transparent ITO electrodes and antiparallel
rubbed polyimide layers for rigid planar surface anchoring of the nematic.
The nematic liquid crystal was a four component mixture (\mbox{\em Mischung 5})
of alkyloxyphenyl-alkyl(oxy)benzoates \\
\def\benz{\unitlength=1pt
\linethickness{0.5pt}
\begin{picture}(21,15)
\put(5.00,14.00){\line(1,0){10.00}}
\put(5.00,-6.00){\line(1,0){10.00}}
\put(0.00,4.00){\line(1,2){5.00}}
\put(0.00,4.00){\line(1,-2){5.00}}
\put(20.00,4.00){\line(-1,-2){5.00}}
\put(20.00,4.00){\line(-1,2){5.00}}
\put(10.00,4.00){\circle{12}}
\end{picture}
}
1O-O6:  22.0\%       \hspace{2cm}
5O-O8:  30.3\%                                       \\
6O-O7:  13.3\%       \hspace{2cm}
6\ -O4: 34.4\%. \\
n(O)-Om= $C_n H_{2n+1}-(O)-\benz-COO-\benz-O-C_m H_{2m+1}$ . 
\medskip

Its nematic range extends from 70.5$^\circ$C to below RT.
\mbox{\em Mischung 5} is well characterized as a nematic standard substance
(e.g. \cite{Schuhmacher,DaAmm}).
The substance has a small negative dielectric anisotropy
$\epsilon_\parallel-\epsilon_\perp$.
All measurements were performed at $30^\circ$C.
The refractive indices at $30^\circ$C are $n_e=1.6315$ and $n_o=1.4935$.
These values have also been used in all simulations.

Our polarizer-analyzer setup is shown in Fig. \ref{setup}, two
different geometries have been applied.
In Fig. \ref{setup}a, the polarizers are along the rubbing direction
such that only the transmission of extraordinary incident light is
observed. In this geometry, the convection roll patterns can be studied
and information on the director tilt profile is obtained.
In the second arrangement shown in Fig. \ref{setup}b, the polarizer
has been rotated perpendicular to the director alignment.
Thus, only ordinary incident light is present, and a contrast modulation
due to refraction in the periodic tilt profile within the 
convection rolls is avoided.
The quarterwave plate at $45^\circ$ to the polarizer discriminates 
domains of right-handed and left-handed director twist. The position
of the analyzer can be chosen either parallel or perpendicular to the
polarizer, yielding equivalent information (intensities are complementary).
The transmission images are recorded with a conventional CCD camera
and digitized for numerical processing.

It would be desirable in the experiment to take both the tilt and twist
sensitive images concurrently, in particular for non-stationary textures.
In practice, however, we had to switch polarizer and $\lambda/4$
plate quickly between individual snapshots of the transmission textures.
Rapidly fluctuating images (timescale $<1s$) can therefore be compared
only in a qualitative way.

\section*{OPTICS}

The simulation of the optical characteristics of EC in the standard 
geometry of Fig. \ref{setup}a has been extensively studied
\cite{DissRasenat,Vistin,Kondo,Richter,Mylnikov,Rasenat}.
The optical pattern is generated by diffraction of the incoming 
extraordinary wave in the periodically modulated refractive index
profile of the deformed liquid crystal.
The transmission texture of roll cells is a periodic stripe pattern
which changes its appearance with the microscopic focus plane.
One observes the optical pattern if the incident light is unpolarized 
or polarized along the director easy axis $\vec n_0$. The convection roll 
textures disappear when the incident light is polarized perpendicular to 
$\vec n_0$.

Optical transmission of a twisted nematic layer with a
phase plate in the geometry of Fig. \ref{setup}b
is influenced by completely different effects.
When the polarizers are perpendicular to $\vec n_0$, 
we deal with the ordinary incident wave which experiences
a constant refractive index independent of director tilt.
The texture of a non-twisted but tilt-modulated director field ($n_y=0$) 
is uniform. The diffraction effects can be neglected.
If the director twists out of the $(x,z)$ tilt plane, the ordinary wave is
guided with the twisted director as long as the {\em Mauguin} condition
holds, irrespective of small tilt deformations.
The exact analysis shows that the polarization state of
incident light is not completely restored.
In general the transmitted light is no longer linearly polarized but
becomes elliptically polarized.
The main axis of the ellipse of the exit beam is rotated out of
the incident polarization direction. Due to the symmetry of the
system, the rotation is reverse for oppositely twisted domains
of the director field, and the effect of the quarterwave plate is 
contrary. 

For a calculation of the optical effects of the director twist, one
can apply several approximative methods. 
A general solution for one-dimensionally deformed nematic layers can be
found from the Berreman \cite{teitler,berreman} and Jones \cite{jones,jones2}
matrix formalism, with and without consideration of internal reflections.
Gerber and Schadt have introduced
an analytical expression which is valid for small angle twist.
Scheffer and Nehring, e.g. have applied a transfer matrix method for the
calculation of the optical characteristics of strongly twisted nematic cells
\cite{scheffer}.
We use this approach and modify the transfer matrix appropriately to describe 
our experimental observations of the chevron textures.
The differences which have to be considered are first that the
twist angle is modulated in the cell plane whereas in the twist 
Fr\'eedericksz transition or in twisted nematic cells, the deformation 
is uniform in the cell plane, and second we have to take the phase plate into
consideration. Finally, there is an additional 
short-wavelength tilt modulation due to the convection rolls which should
affect the extraordinary incident wave, but also the ordinary incident wave
at high twist deformations.

The computation of the optical transmission can be performed 
when some approximations are introduced. 
We demonstrate for the geometry of Fig. \ref{setup}b how periodical
twist modulations in the sample plane influence the transmission texture.
We begin with a director field in the sample which be
homogeneous in the $(x,y)$ plane at least on distances large compared to
the wavelength, such that plane waves can be assumed.
We establish $2\times 2$ optical propagation matrix
\begin{equation}
\label{propagator}
\bar P = 
\bar P_A \times \bar P_\lambda \times \bar P_{LC} \times \bar P_P
\end{equation}

from the polarizer matrix $\bar P_P$, 
the propagator matrix in the nematic layer $\bar P_{LC}$, 
the propagator matrix of the $\lambda/4$ plate $\bar P_\lambda$,
and that of the analyzer $\bar P_A$,
$$
\bar P_P=\frac{1}{2}\left[
\begin{array}{cc}
 1+\cos 2\alpha_p       & \sin 2\alpha_p \\
 \sin 2\alpha_p         & 1-\cos 2\alpha_p
\end{array}
\right]    ,
$$

$$
\bar P_\lambda=\left[
\begin{array}{cc}
 \cos^2\alpha_\lambda + i \sin^2\alpha_\lambda  & 
 \sin\alpha_\lambda \cos\alpha_\lambda (1-i) \\
 \sin\alpha_\lambda \cos\alpha_\lambda (1-i)    & 
 \sin^2\alpha_\lambda + i \cos^2\alpha_\lambda
\end{array}
\right]  ,
$$
$$
\bar P_A=\frac{1}{2}\left[
\begin{array}{cc}
 1+\cos 2\alpha_a      & \sin2\alpha_a \\
 \sin 2\alpha_a        & 1-\cos 2\alpha_a
\end{array}
\right]   ,
$$
and $\alpha_p, \alpha_\lambda, \alpha_a$ 
are the rotation angles of polarizer, phase plate 
and analyzer with respect to the director easy axis.


The propagation matrix in the nematic layer can be calculated 
when the nematic is treated as a stratified medium. It is
described as a stack of sufficiently thin layers in which the nematic 
material can be considered homogeneous and the propagation matrix 
in each slice $n$ is 
$$
\bar P_n=\left[
\begin{array}{cc}
\exp(2\pi i n_e d_n/\lambda)& 0     \\
0            & \exp(2\pi i n_o d_n/\lambda)
 \end{array}
 \right]
$$
in the local principal axis frame of the director system,
with the refractive indices $n_{e,o}$ of the nematic, the vacuum
wavelength $\lambda$ and the thickness $d_n$ of the slices.
In the laboratory system it reads
$$
\bar P_n=\exp(i\psi) \left[
\begin{array}{cc}
 \cos\gamma + i \sin\gamma \cos 2\phi &      \sin\gamma \sin 2\phi \\
    i \sin\gamma \sin 2\phi   &   \cos\gamma - i \sin\gamma \cos 2\phi
 \end{array}
 \right] 
$$
for a slice with the director in the angle $\phi$ respective to the $x$ axis,
with 
$
\gamma=\pi (n_e-n_o) d_n/\lambda,
$
and
$
\psi=\pi (n_e+n_o) d_n/\lambda.
$
The total propagator $\bar P_{LC}$ is the product 
$$
\bar P_{LC} =\prod_{n=1}^{N} \bar P_n
$$ 
of the matrices of the individual layers.
This formalism neglects internal reflections in the LC layer and is 
thus only an approximation, although a sufficiently good one, to the 
exact solution which could be found by Berremans $4\times 4$ formalism. 
The elements of the resulting matrix $\bar P$ (Eq. \ref{propagator})
give the ratios of input vs. output electric field vectors 
$E_x^{(in)}$, $E_y^{(in)}$, $E_x^{(out)}$, $E_y^{(out)}$. 
The ratios of transmitted vs. incident intensities
are $|P_{ij}|^2$ for the four different combinations of input and output 
polarization, and
$$
I= \sum_{i,j=1}^2 |P_{ij}|^2
$$
is the total transmission coefficient of unpolarized light.

If the director twist deformation $\phi(z)$ in the sample is known, the
relation between director field and optical 
transmission is calculated straightforward.
Under the assumption that the director deformation is a pure
twist ground mode,
$\vec n=(\cos(\phi(z)),\sin(\phi(z)),0)$
with $\phi(z)=\phi_m \cos(z\pi/d)$, the transmission intensity as
a function of the maximum twist angle (in the middle of the cell)
is shown in Fig. \ref{rechnung1}. A cell thickness of $50.5\mu$m and
optical parameters of \mbox{\em Mischung 5} have been assumed, the results
are shown for three different optical wavelengths. 
The transmission intensity changes quite
linearly with the maximum twist angle when the twist deformation is small, 
and the curve is antisymmetric with respect to $\phi_m=0$. 
Left-handed or right-handed twist lead to opposite changes
in the optical transmission. 
Strictly, the calculations are correct for an in-plane uniform sample,
but when the twist domains are large compared to the optical wavelengths, 
refraction can be neglected and the results remain valid.  
Oppositely twisted domains appear in the optical pattern as regions of
different brightness, a contrast modulation in the transmission
texture is expected.
In the experiment, comparison of absolute intensities is quite difficult.
Therefore we define the contrast 
$C=2( I_+ - I_-)/( I_+ + I_-)$ from the difference of the transmission 
intensity extrema $I_+, I_-$ of oppositely twisted domains in a sample with
periodic twist modulations. This optical contrast is linearly related
to the twist amplitude at least for small twist ($<45^\circ$).
Of course, at high twist angles the relation between $\phi_m$ and
the transmitted intensity is no longer linear. The director 
twist deformation along $z$ will no longer be sinusoidal, other model
director fields have to be assumed. 
Therefore, we restrict the quantitative comparison to voltages near the
onset of the chevron texture.

The calculations above have  been performed for three different wavelengths
only, and the transmission appears to be strongly wavelength dependent.
Fig. \ref{rechnung3} shows the linear slope of the transmission curve
($C/\phi_m$ in the vicinity of $\phi_m=0$) calculated numerically 
for the complete optical wavelengths band. It shows that knowledge of the 
wavelength (actually the wavelength / cell thickness ratio )
is critical when one is interested in quantitative results.
When the sample is illuminated with white light and observed with a B/W 
camera as in our experiments, 
it is a sound approximation to use an average slope
of 0.0027/deg for the $C$ vs. $\phi_m$ ratio
to establish an approximate quantitative relation between the twist 
deformation in the sample and measured optical contrast.

Fig. \ref{rechnung2} shows the optical transmission as a function
of director twist when no phase plate is present 
(the conventional setup of Fig. \ref{setup}a).
The function is now symmetric with respect to positive and negative 
twist and the optical contrast is considerably smaller. 
Only at very high twist deformations, one obtains a substantial optical
modulation, but now between non-twisted ($\phi_m=0$) and twisted regions,
that is with 1/2 wavelength of the twist pattern.

Two other arrangements of the polarizers can be described easily.
If the analyzer in \ref{setup}b is rotated to $90^\circ$, 
the optical transmission is
complementary to that in Figs. \ref{rechnung1} and \ref{rechnung2}.
If the polarizer is turned parallel to the director easy axis,
the short wavelength modulation due to the periodic tilt deformation
(dielectric convection rolls) will be effective from the beginning and
overlaps the interference effect (see experiments).

The more the director field is twisted, the larger is the 
contribution of the extraordinary wave to the transmission image in the
geometry 1b.  In that case, the periodic tilt deformation 
and the diffraction effects of the convection rolls come into play again.

\section*{EXPERIMENTAL RESULTS}

Figs. \ref{image1}-\ref{image3} present a collection of transmission
textures under different experimental conditions.
All voltages given are scaled with the threshold voltage $U_c=67.4V$
of the onset of dielectric rolls to the reduced quantity $r=U/U_c$.
Fig. \ref{image1} shows the first instability above threshold, 
normal dielectric rolls, observed in standard
geometry \ref{setup}a. Figs. \ref{image2},\ref{image3} depict the 
development of the chevron pattern at $r=1.159$ and $r=1.605$, respectively.
The first image (a) is observed with
setup \ref{setup}a, the second (b) shows the texture
with a quarterwave plate inserted at $45^\circ$, the third (c)
corresponds to the setup in Fig. \ref{setup}b and the last image (d)
is the texture under crossed polarizers without phase plate.
The director field in normal dielectric rolls (Fig. \ref{image1}) 
at $r=1.005$ is 
planar in $(x,z)$, therefore the images in other geometries than setup 
\ref{setup}a are trivial and have not been shown there.

Fig. \ref{image2}, 
shows a chevron texture at small voltages where the
periodic twist deflection already leads to transmission intensity
modulations when the phase plate is inserted (\ref{image2}b). 
The oppositely twisted domains are visible, in particular in the ordinary
light where the roll pattern disappears (\ref{image2}c).
However, a completely uniform dark image is retained under crossed polarizers
when the phase plate is removed (\ref{image2}d). The
director twist is too weak to produce substantial contrast between 
twisted and non-twisted regions.

The comparison of the twist structure visualized in images 
(b,c) with
the corresponding convective rolls pattern (a,b) shows exact
agreement of the spatial twist angle modulation and roll pattern. 
It has to be mentioned also that the director and wave vector
modulation have the same phase, i.e. the director is deflected clockwise where
the wave vector is deflected clockwise and vice versa.

Fig. \ref{image3} presents a chevron texture at high fields where
the contrast modulation in geometry \ref{setup}b is considerable.
Even with crossed polarizers and no phase plate, a texture modulation 
between non-twisted and strongly twisted regions is seen (\ref{image3}d).
(Note, however that the contrast of the image presented in Fig. \ref{image3}d 
has been enhanced very much.) 
Fig. \ref{schnitte} visualizes a cross section of the intensity taken normal
to the chevrons. The upper curves correspond to Figs. \ref{image2}c (solid
line) and \ref{image3}c (dashed) resp., 
they give the intensity modulation with setup \ref{setup}b
(linear effect). One acknowledges that the spatial modulation is
approximately sinusoidal and the modulation amplitude increases with $r$.
The lower curves correspond to profiles under crossed polarizers 
(quadratic effect). The observed modulation wavelength is half that of the 
linear effect, 
the amplitude is much weaker, therefore this effect is only detectable at 
high $r$.

A more quantitative analysis is presented in Figs.
\ref{wavenumber}-\ref{contrast}.
The wavelength of the convection rolls  is shown in 
Fig. \ref{wavenumber}. It has been extracted from the 2D Fourier transforms
of the chevron textures. 
The wavevectors of the rolls change their orientations away from the 
$x$ axis with increasing voltages, but their absolute value (wavenumber) 
is almost unchanged over the whole voltage range. 
Fig. \ref{winkel} shows the angles between $\vec n_0\parallel x$ and the wave
vectors of the rolls (maximum angles in the middle of the chevron stripes), 
obtained also by Fourier transform of the images.
The transition from normal to inclined rolls is clearly indicated \cite{obl}. 
Left-handed and right-handed domains appear nearly symmetrically.
Fig. \ref{contrast} depicts the corresponding contrast $C$
between the brightness $I_+$ and $I_-$, resp., of oppositely twisted domains.
The increase in contrast ratio is an empirical measure of the increasing
director twist with higher voltage. The exact relation between
the twist angle and contrast, however, is quite complex as has been  shown 
in Figs.  \ref{rechnung1} and \ref{rechnung3}.
When we assume a simple $\phi(z)=\phi_m\cos(z)$ dependence of the twist angle 
in the cell and an approximately linear relation between contrast and maximum 
twist $\phi_m$ with the factor 0.0027/deg  derived in the previous section, 
we can estimate the twist angle modulations. 
The corresponding scale is shown at the right of Fig. \ref{contrast}.
We note, however, that this linear relation bases upon the simulation for
small twist angles (see above). The higher contrast measured at high
electric fields can be related to the twist angle only qualitatively.
The figure shows that the optical contrast and correspondingly the director
twist follow roughly the characteristics of the inclination
angle of the wave vector.

Finally, we investigate the evolution of the texture after the electric 
field has been switched off. The relaxation behaviour will help us to
extract information on the elastic modes involved in the $z$-profile of
the twist deformation.
Fig. \ref{offsw}a depicts a spatio-temporal image taken with a line scan
CCD device. Each horizontal line represents the transmission profile
in a cross section taken along the director easy axis $\vec n_0$, 
the vertical axis 
gives the time coordinate $t$ running from top to bottom. 

The image \ref{offsw}a was taken with polarizers along $x$ and a
phase plate, the observed refraction pattern due to the modulated director 
tilt in the convective rolls and the long-range twist pattern are 
superimposed. One can see that while the electric AC field is effective
the pattern is strongly modulated in time because the director tilt
oscillates with the AC frequency. In addition, one can see five spatial
periods of the long-wavelength tilt modulation which correspond to a cut
normal to the chevron superstructure
(as in Fig. \ref{schnitte}, upper curves). After the field is switched off,
the tilt modulation decays immediately, within several milliseconds. 
The moment the electric field was switched off is clearly indicated  
in the image by the abrupt stop of the high-frequency modulations, 
the decay time is shorter than the temporal resolution 
chosen here. This was expected as the driving flow field
breaks down instantaneously, and the short wavelength roll pattern 
is rapidly destroyed by elastic restoring forces.
On the other hand, the twist modulation persists for a long time of several 
seconds after the field has been switched off. The elastic forces which
destroy the long wavelength twist pattern are much smaller than those
effective in the destruction of the short wavelength convection roll texture.
The corresponding decay times are therefore much larger.
Fig. \ref{offsw}b shows the decay of the amplitude of the 
long-wavelength modulation seen in Fig. \ref{offsw}a. We have depicted the
absolute value of the respective coefficient in the Fourier transform.
The solid line is a fit to a purely monoexponential decay with decay time 
$\tau=2.5s$.
A rough theoretical estimate of the expected relaxation time  
for a ground mode sine deformation in $z$ can be obtained from 
the elastic twist and bend terms involved and the rotational viscosity
$\gamma_1$
$$
\tau \approx \gamma_1 (K_{22} q_z^2 + K_{33} q_x^2)^{-1}
$$
with $q_z=\pi/50\mu m, q_x=2\pi/60\mu m$ and the viscoelastic parameters 
$\gamma_1=0.36 Pa s, K_{33}=13.8 pN, K_{22}=6.0pN$ taken from independent
magneto-optic measurements of the dynamic and static Fr\'eedericks 
transitions, the calculated decay time is 2.1s.
In view of the rough approximations used in this calculation, 
this is in quite reasonable agreement with the experimental decay rate.

\section*{DISCUSSION AND SUMMARY}

We have shown that the optical investigation of chevron patterns with
different microscopic techniques reveals new details of the
director field in dielectric dissipative convection structures.

At voltages slightly above the onset of dielectric rolls, we observe
stationary normal rolls. The relation between optical texture and 
director field in these rolls has been shown in a previous paper \cite{prep}.
At increasing voltage, the director field twists slightly out of the 
$(x,z)$ tilt plane, accompanied by a corresponding rotation of the wave vector
of the rolls out of the $x$ axis direction. Both processes set in at
approximately the same critical voltage. The director twist and 
wave vector modulation follow roughly the same characteristics when
the voltage is increased, and both deflections are in phase, such that
the structures roughly remain normal rolls on the local scale.
The absolute angle of the wave vector modulation can be extracted from
the transmission image. The exact director twist angle was estimated from
a comparison of the experimental optical images with numerical calculations 
of the transmission intensity.

The dynamics of the tilt and twist deformations are quite different.
As the wavelength of the twist structure is large compared to that of the
convection rolls (it corresponds to
the periodicity of the chevrons), its elastic restoring forces are very
weak, and the twist mode relaxes very slowly.
The twist modulation is therefore approximately constant during the cycles 
of the driving field. 
This was established in time resolved measurements by means of a fast
line scan camera. The tilt modulation of the director
in the convection rolls is dynamically fast. It alternates its sign during
the cycles of the driving field. This well known theoretical result was
confirmed in a previous experiment \cite{prep}.

From the nearly monoexponential decay of the twist deformation after the
driving field is switched off (Figs. \ref{offsw}a,b), 
we conclude that the $z$-deformation is 
basically restricted to the sine ground mode $\propto \sin \pi z/d$
at least near the onset of chevron formation. Higher twist deformation modes 
should decay much faster and the total decay curve should be non-exponential 
when more than the ground mode contribute to the $z$-profile substantially. 

In summary, the experiments provide information on the relation between
director and wave vector structures in chevrons observed in a planar cell.
In the theoretical paper by Rossberg \cite{Rossberg}, simulations showed
that the chevron formation is a consequence of the interactions between
the in-plane director twist and wave vector fields.  
It has been argued that this mechanism is general for systems with
weakly broken isotropic symmetry, but is quite difficult to
verify in planar nematic cells. 
Our experimental setup provides exactly this test.
The results presented are in accordance with the qualitative
predictions by Rossberg \cite{Rossberg} obtained from the simulations.

In the conduction regime, director 
twist modulations are observed as well when the driving electric field is
sufficiently above the threshold field for conductive rolls.
We find that these twist modulations set in approximately at the same 
field strengths where the stationary normal or oblique rolls start to
fluctuate. The study of this effect in the conduction regime
may contribute to the understanding of the mechanisms leading to abnormal 
and fluctuating rolls
and will be  an interesting future task.
 
The authors are very indebted to L. Kramer for many useful hints and
critical discussions.
This study was supported by the DFG with grant STA 425/3-1.

\small
\bibliographystyle{plain}

\begin{thebibliography}{10}
\bibitem{obl}
We do not use the term oblique rolls here because the director still
seems to be rather normal to the convection rolls.
\end{thebibliography}

\begin{figure}
\caption{
\label{setup}
Experimental geometry of the cell, polarizers and $\lambda/4$ plate.
The rubbing direction is $x$, $z$ is normal to the cell plane.
{\bf (a)} setup for the observation of the director tilt profile
{\bf (b)} setup sensitive to the director twist profile. 
{\bf (c)} Schematic presentation of the cell with the convection rolls in 
the chevron texture (top) and the corresponding director twist in the sample 
(bottom).
}
\end{figure}

\begin{figure}
\caption{
\label{rechnung1}
Calculated optical transmission for the setup in 
Fig. \protect{\ref{setup}}b, with $d=50.5\mu{}m$,
$n_e=1.6315$, $n_o=1.4935$, and a sine dependence of the director 
twist $\phi(z)=\phi_m\sin\pi z/d$.
The wavelengths are 500nm (---), 550nm (- - -) and 600nm (. . .).
Transmission is normalized to the total incident light passing the polarizer.
}
\end{figure}

\begin{figure}
\caption{
\label{rechnung2}
Same as in Fig. \protect{\ref{rechnung1}} but with setup of
Fig. \protect{\ref{setup}}a. Note the change in the vertical scale.
}
\end{figure}

\begin{figure}
\caption{
\label{rechnung3}
Calculated linear slope of the contrast C vs. $\phi_m$ (per degree) for 
wavelengths in the optical spectrum, 
sample parameters as in Fig. \protect{\ref{rechnung1}}.
}
\end{figure}

\begin{figure}
\caption{
\label{image1}
Transmission texture ($200\mu m\times 200\mu m$)
of the normal dielectric rolls
of \mbox{\em Mischung 5} at U=67.7V ($r=U/U_c=1.005$).
Sample parameters d=$50.5\mu$m, $f=50Hz$, $T=30^\circ$,
$U_c=67.4V$.
}
\end{figure}

\begin{figure}
\caption{
\label{image2}
Transmission textures ($200\mu m\times 200\mu m$) of chevrons
(at $r=1.159$)
under different orientations of the polarizers and quarterwave plate,
sample parameters as in Fig.  \protect{\ref{image1}}. 
(a) conventional technique (polarizers along $x$).
(b) quarterwave plate inserted at $45^\circ$ orientation, 
one sees the convection rolls and the superimposed twist effect,
(c) polarizer in $90^\circ$ orientation to $x$. Now, only
the twist pattern remains visible, the tilt pattern is no longer
observed, the intensity modulation is complementary,
(d) crossed polarizers, no modulation is seen.
}
\end{figure}

\begin{figure}
\caption{
\label{image3}
Images as in Fig. \protect{\ref{image2}} for $r=1.604$.
At this field strength, 
the intensity modulation between non-twisted and twisted regions
becomes visible even under crossed polarizers when the contrast is
sufficiently enhanced by digital processing (d).
}
\end{figure}

\begin{figure} 
\caption{ 
\label{schnitte}
Upper curves: Intensity cross sections of images \protect{\ref{image2}}c 
(solid) and \protect{\ref{image3}}c (dashed). 
The modulation corresponds to the chevrons period.
For comparison, the lower
curves show the corresponding cross sections of images 
\protect{\ref{image2}}d (solid) and \protect{\ref{image3}}d (dashed).
A contrast modulation with half wavelength is indicated in the high 
voltage texture but not visible in the low voltage texture.
}
\end{figure}

\begin{figure}
\caption{
\label{wavenumber}
Experimentally determined wavelength of the convection rolls in the chevron 
texture in dependence upon the reduced driving voltage. $\lambda_0$ gives the
periodicity of the optical image which is half the wavelength of the 
director tilt pattern
(sample parameters as in Fig.  \protect{\ref{image1}}). 
}
\end{figure}

\begin{figure}
\caption{
\label{winkel}
The angle $\theta$ between the wave vector of the convection rolls
and the director easy axis as a function of the reduced voltage
(sample parameters as in Fig.  \protect{\ref{image1}}). The two branches
correspond to the oppositely inclined chevron domains.
}
\end{figure}

\begin{figure}
\caption{
\label{contrast}
Experimentally determined optical contrast C between regions of maximum 
and minimum transmission intensity (opposite director twist) as a 
function of the reduced voltage
(sample parameters as in Fig.  \protect{\ref{image1}}).
Note linear scale given at the right hand side is reliable only
to approximately $45^\circ$, see text.
}
\end{figure}

\begin{figure}
\caption{
\label{offsw}
a) Evolution of the optical pattern after the electric field is switched off 
from $r=1.162$.
The horizontal axis represents a cross section of the transmission image 
along the director easy axis, the vertical coordinate is the time axis (1.6s).
The moment the field is switched off is clearly marked by the end of the
fast temporal oscillations, the slow decay of the twist deformation is
visible.
b) Time dependence of the twist modulation amplitude calculated from the
spatial Fourier transforms. The solid line is a fit to an exponential decay
with time constant 2.5s.
}
\end{figure}

\pagebreak
\large\bf

\hspace*{-2cm}\psfig{figure=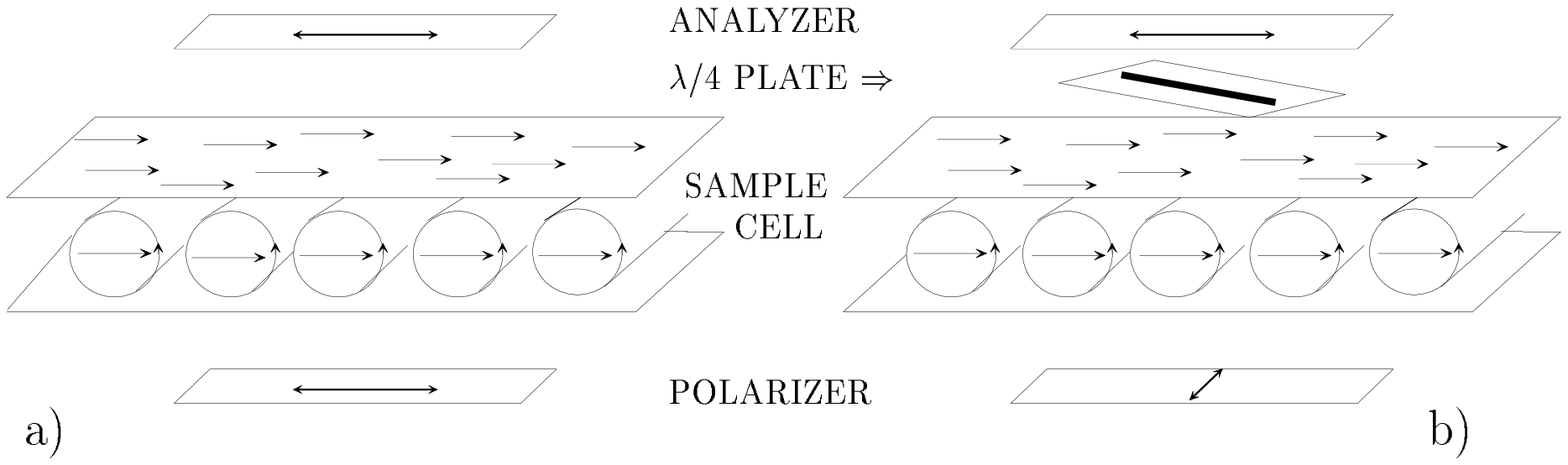}
\vfill
Fig. 1a,b
\pagebreak

\psfig{figure=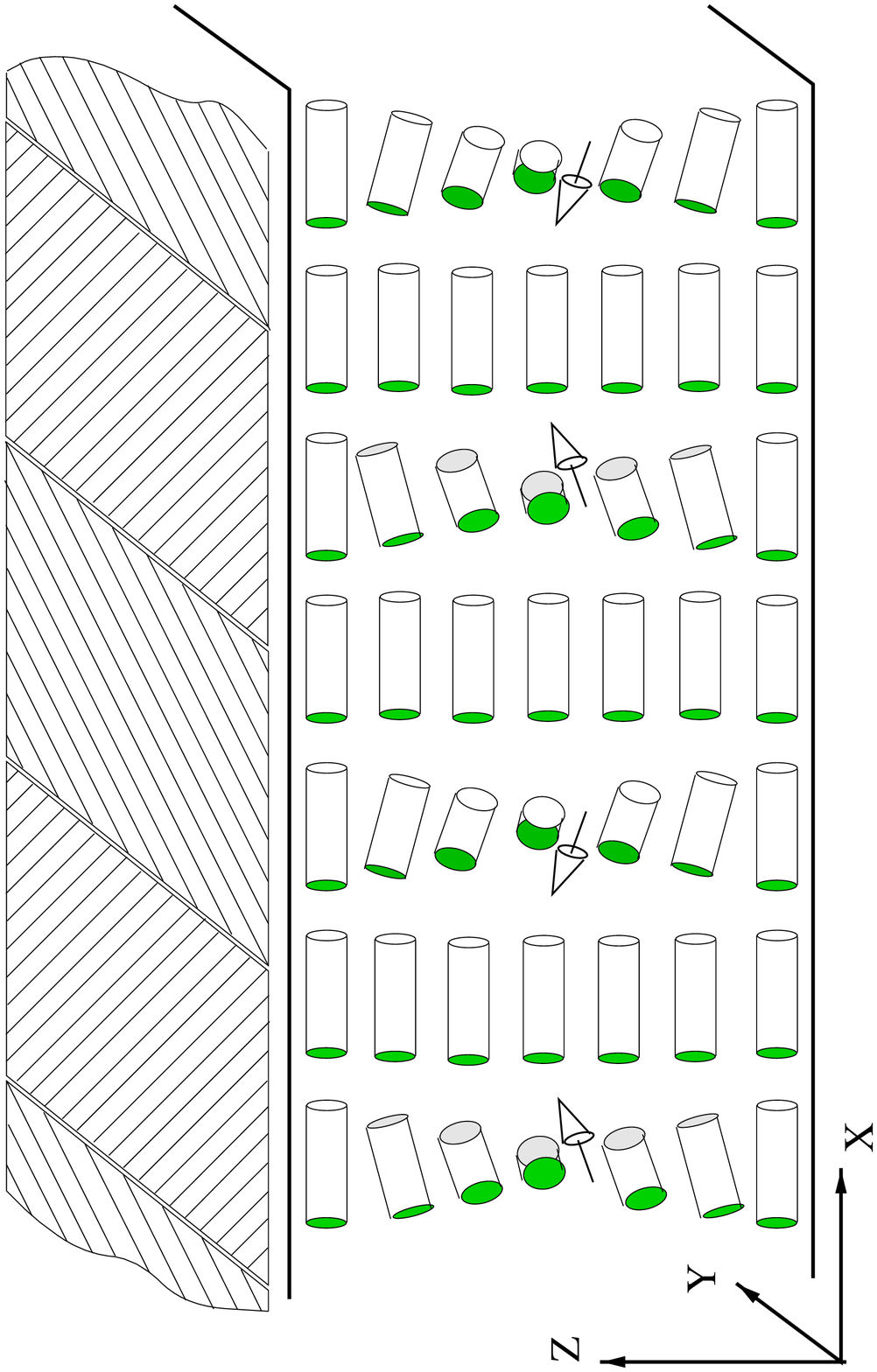,height=20cm}
\vfill
Fig. 1c
\pagebreak

\psfig{figure=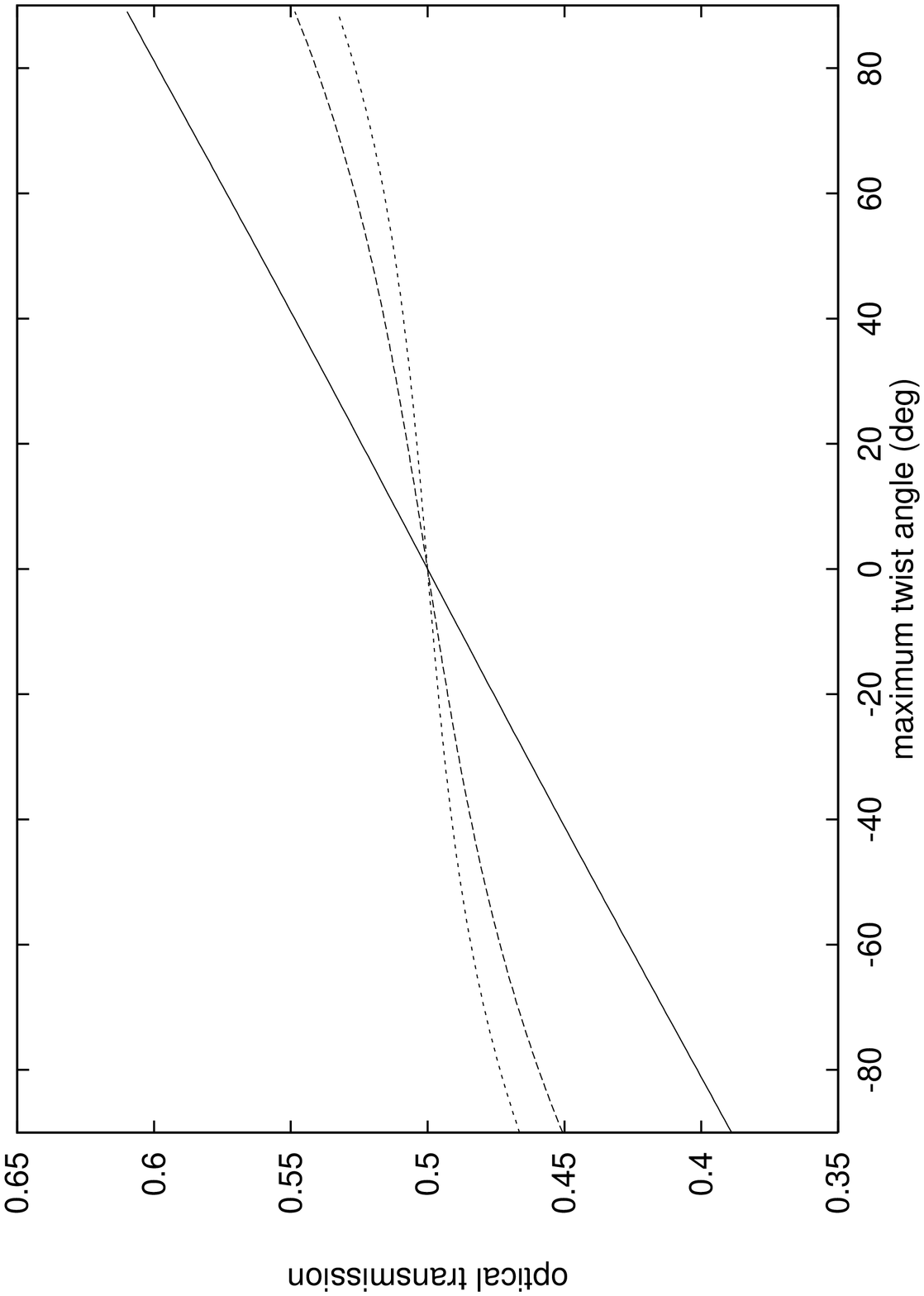,height=20cm}
\vfill
Fig. 2
\pagebreak

\psfig{figure=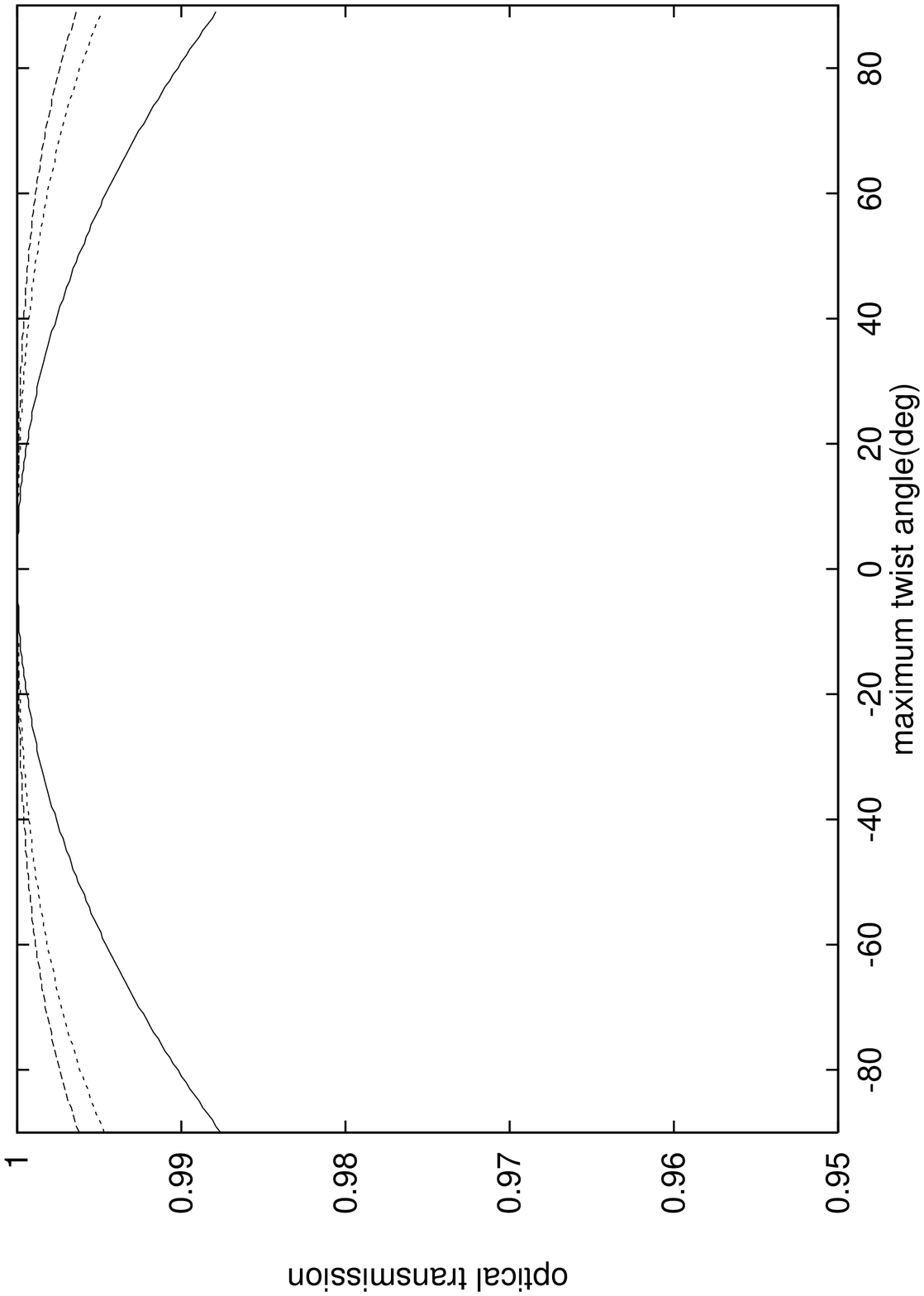,height=20cm}
\vfill
Fig. 3
\pagebreak

\psfig{figure=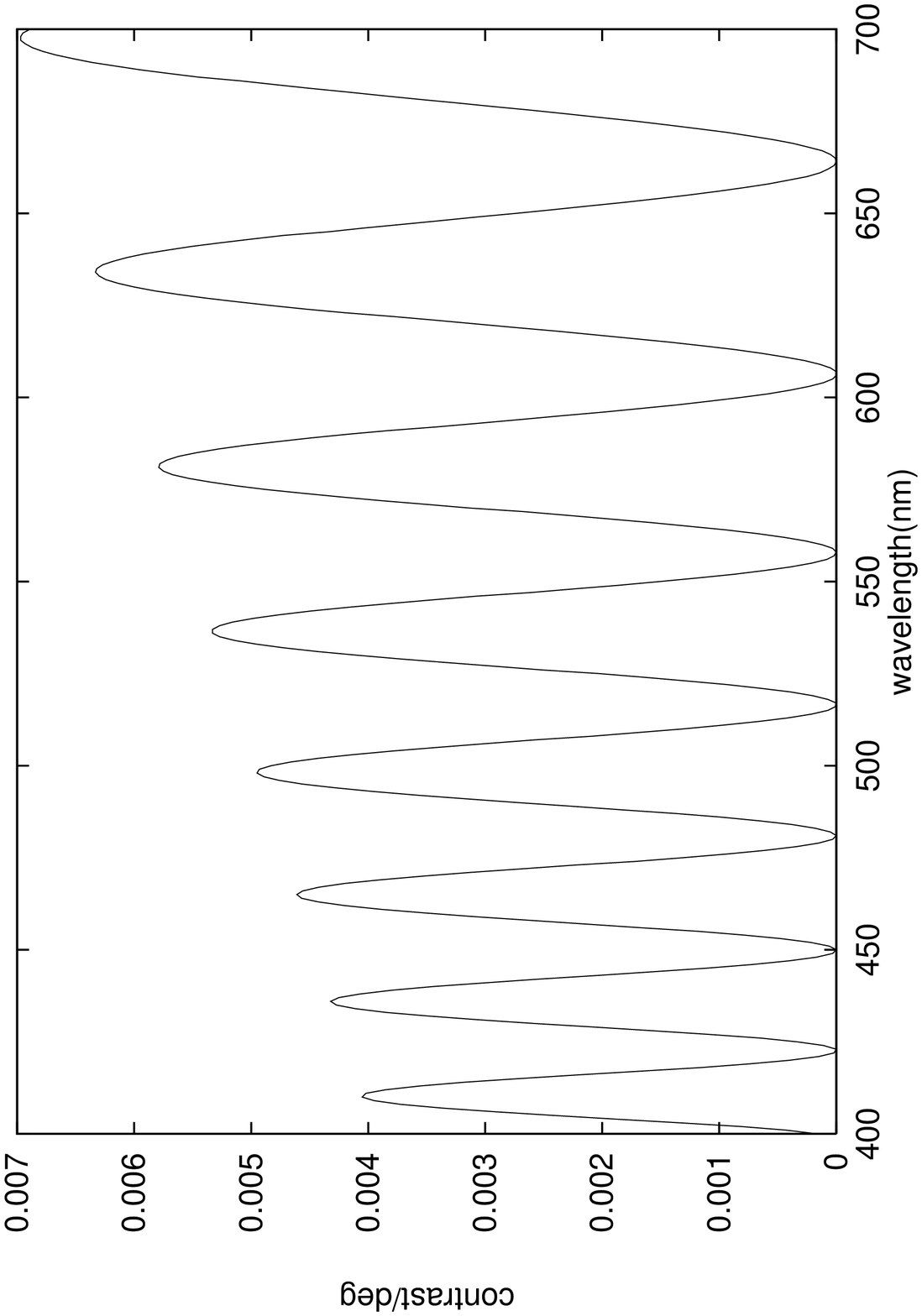,height=20cm}
\vfill
Fig. 4
\pagebreak

\psfig{figure=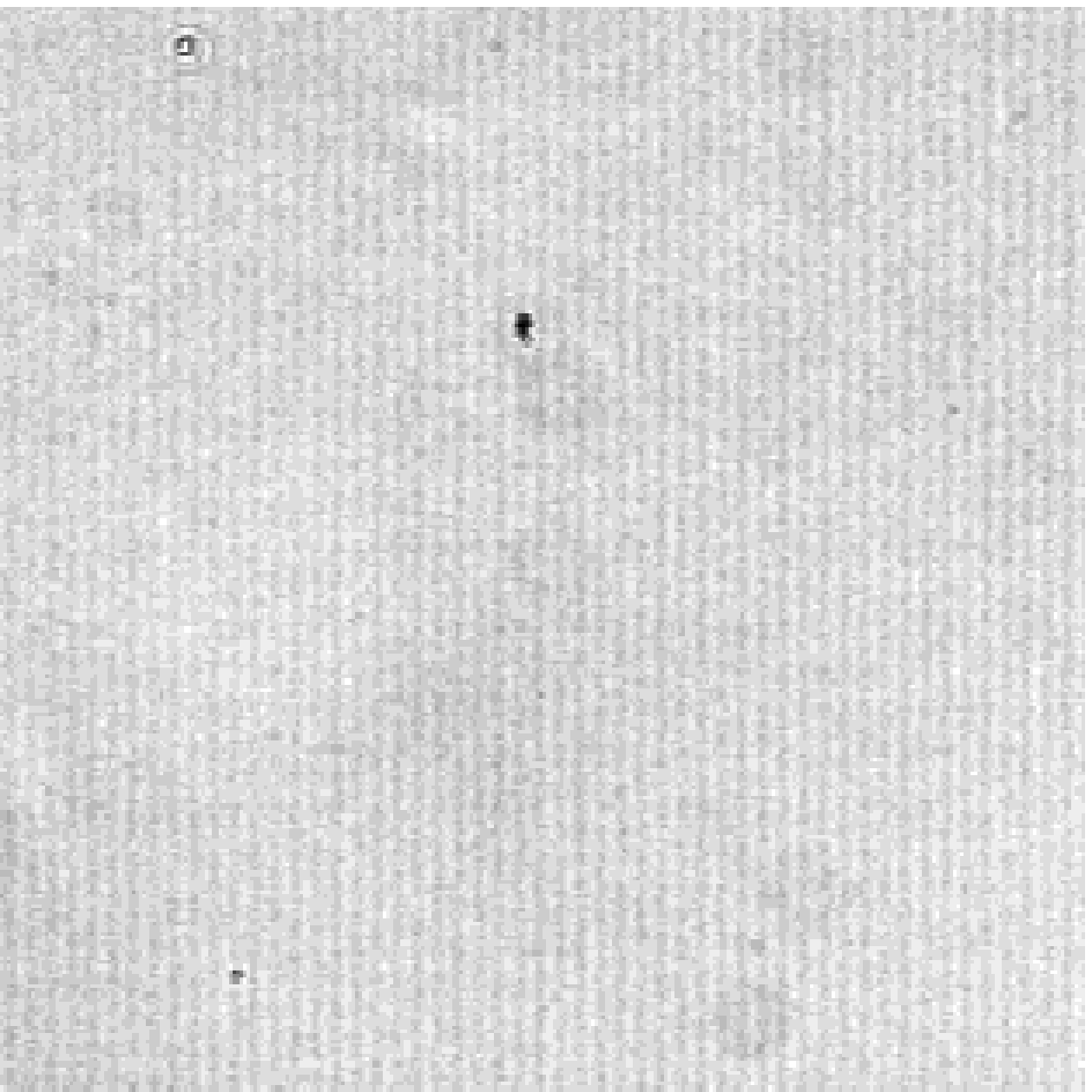,height=10cm}
\vfill
Fig. 5
\pagebreak

a)\psfig{figure=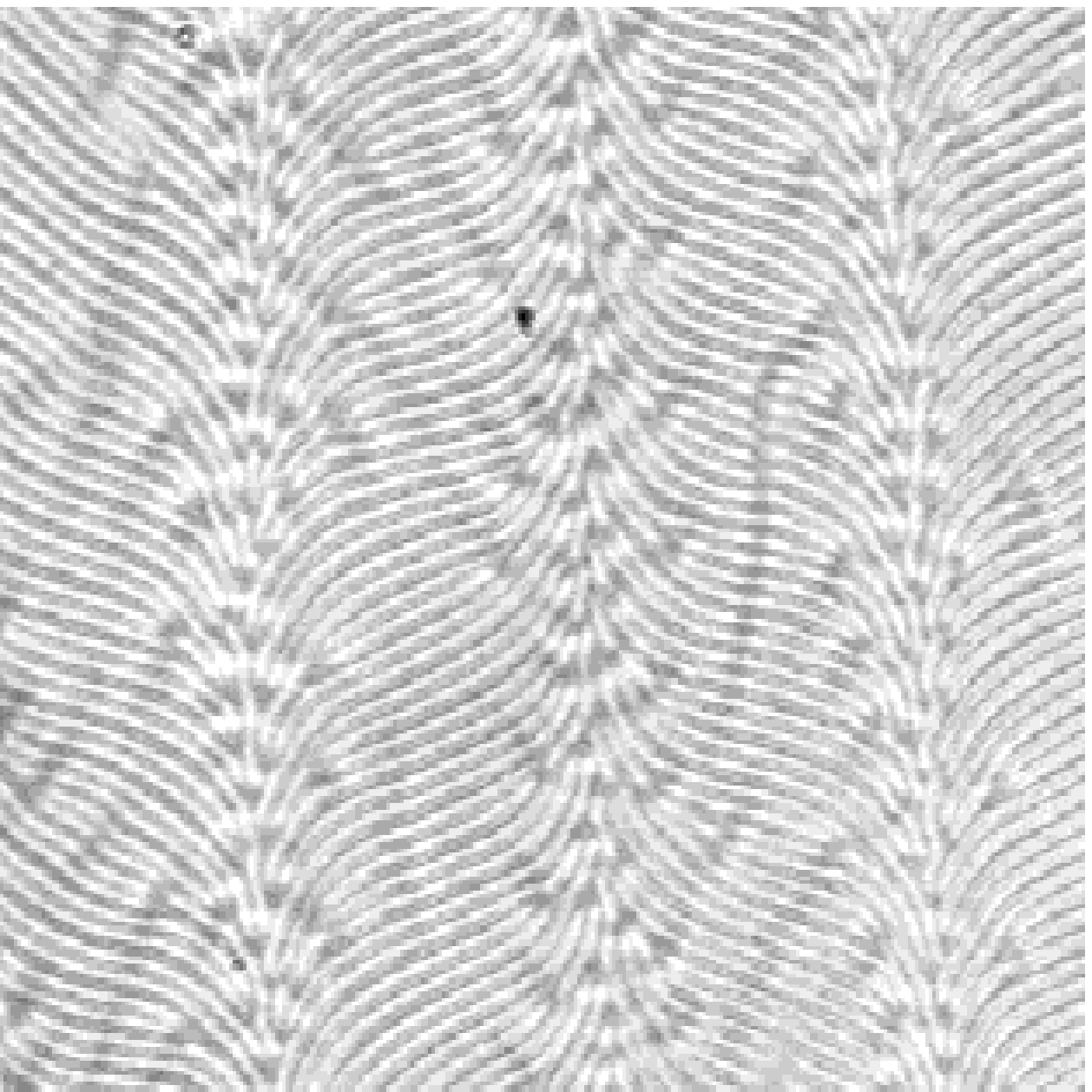,height=10cm}

b)\psfig{figure=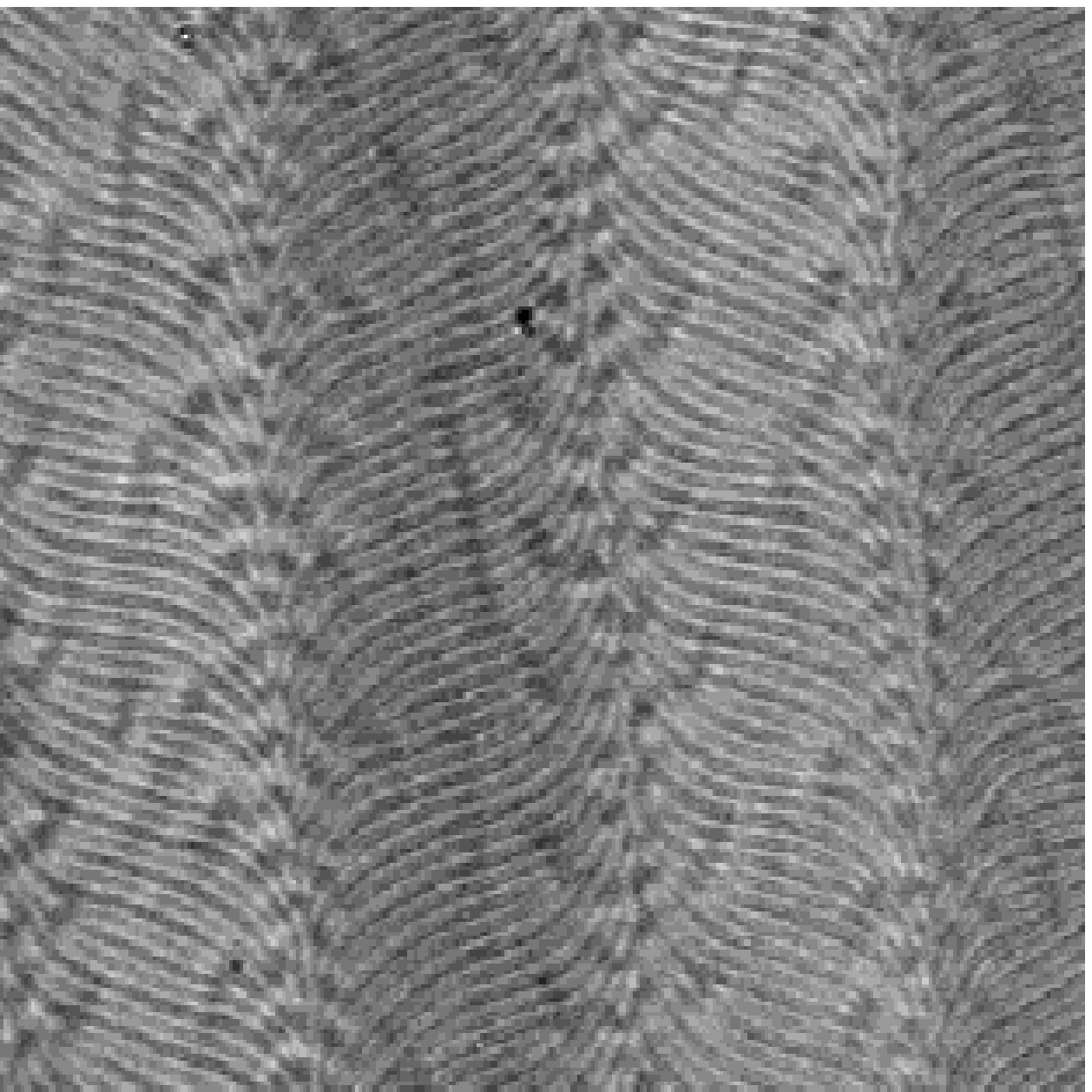,height=10cm}
\vfill
Fig. 6
\pagebreak

c)\psfig{figure=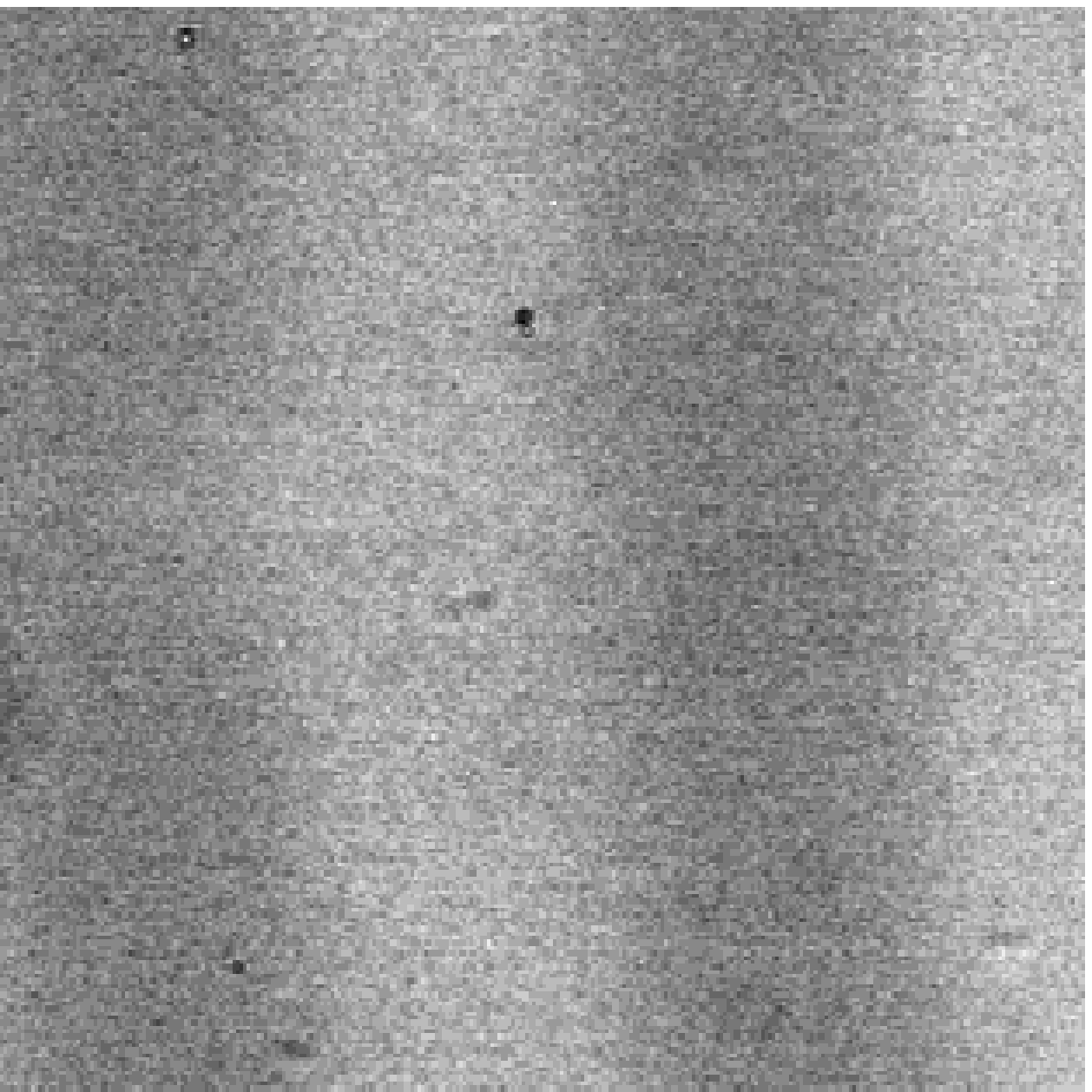,height=10cm}

d)\psfig{figure=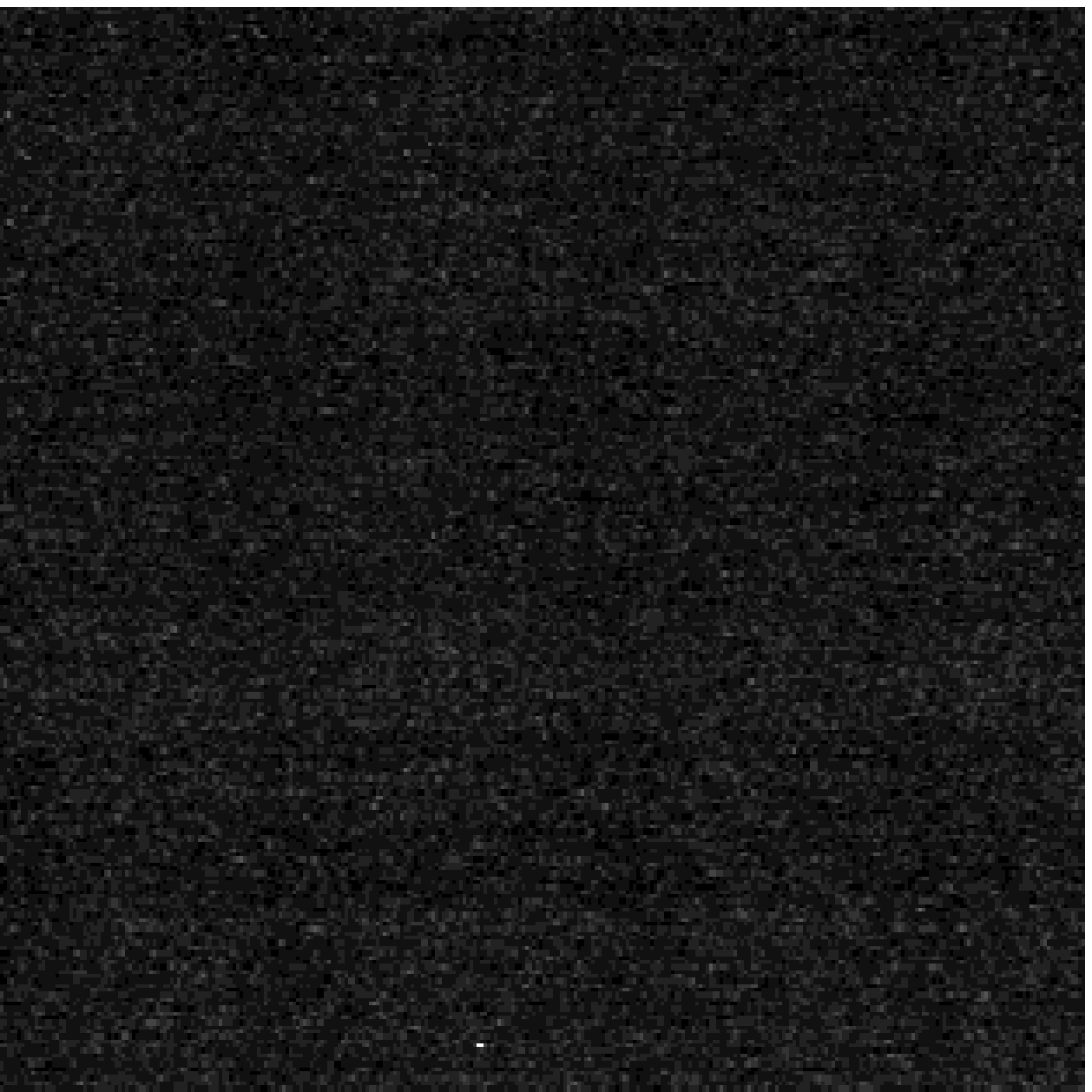,height=10cm}

\vfill
Fig. 6
\pagebreak

a)\psfig{figure=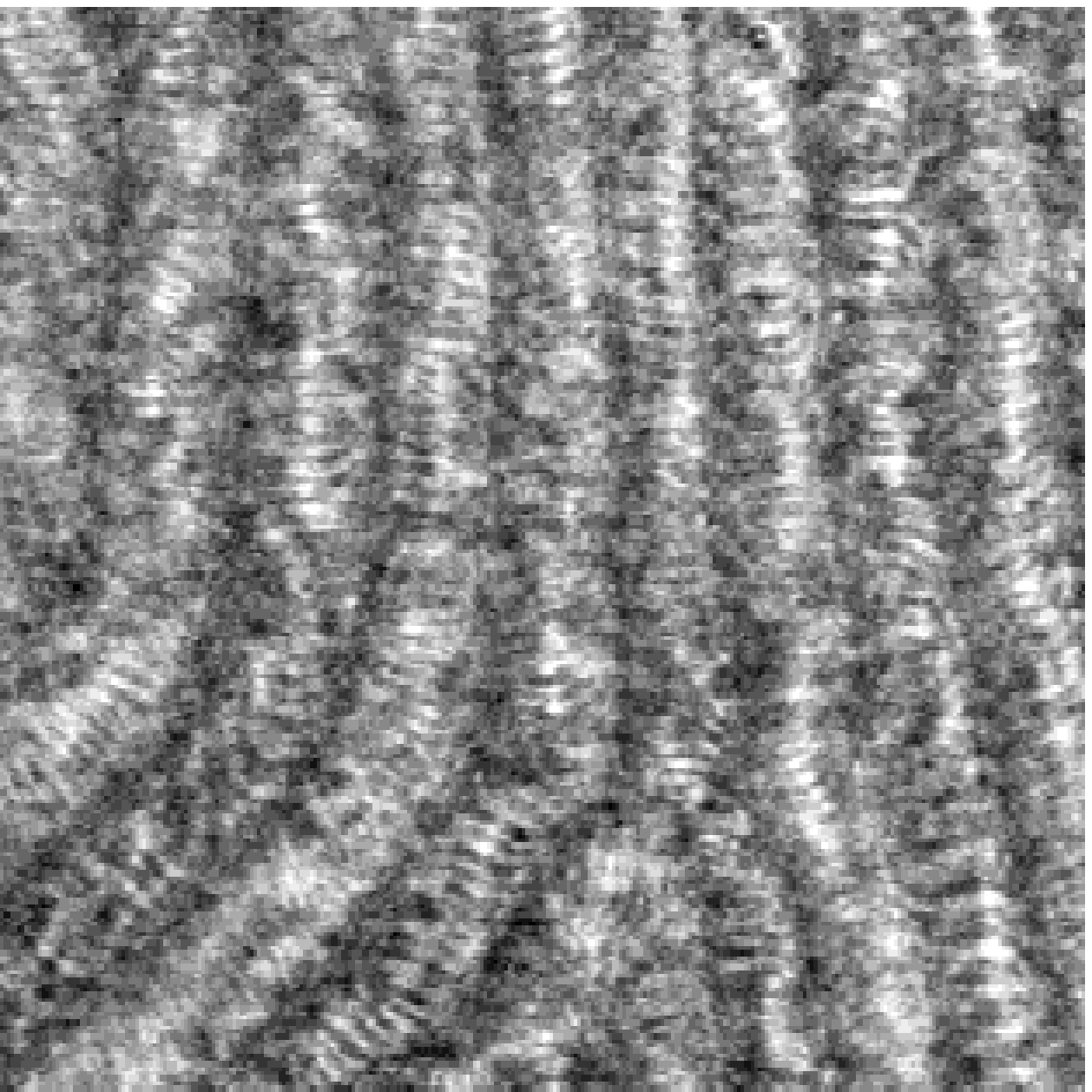,height=10cm}

b)\psfig{figure=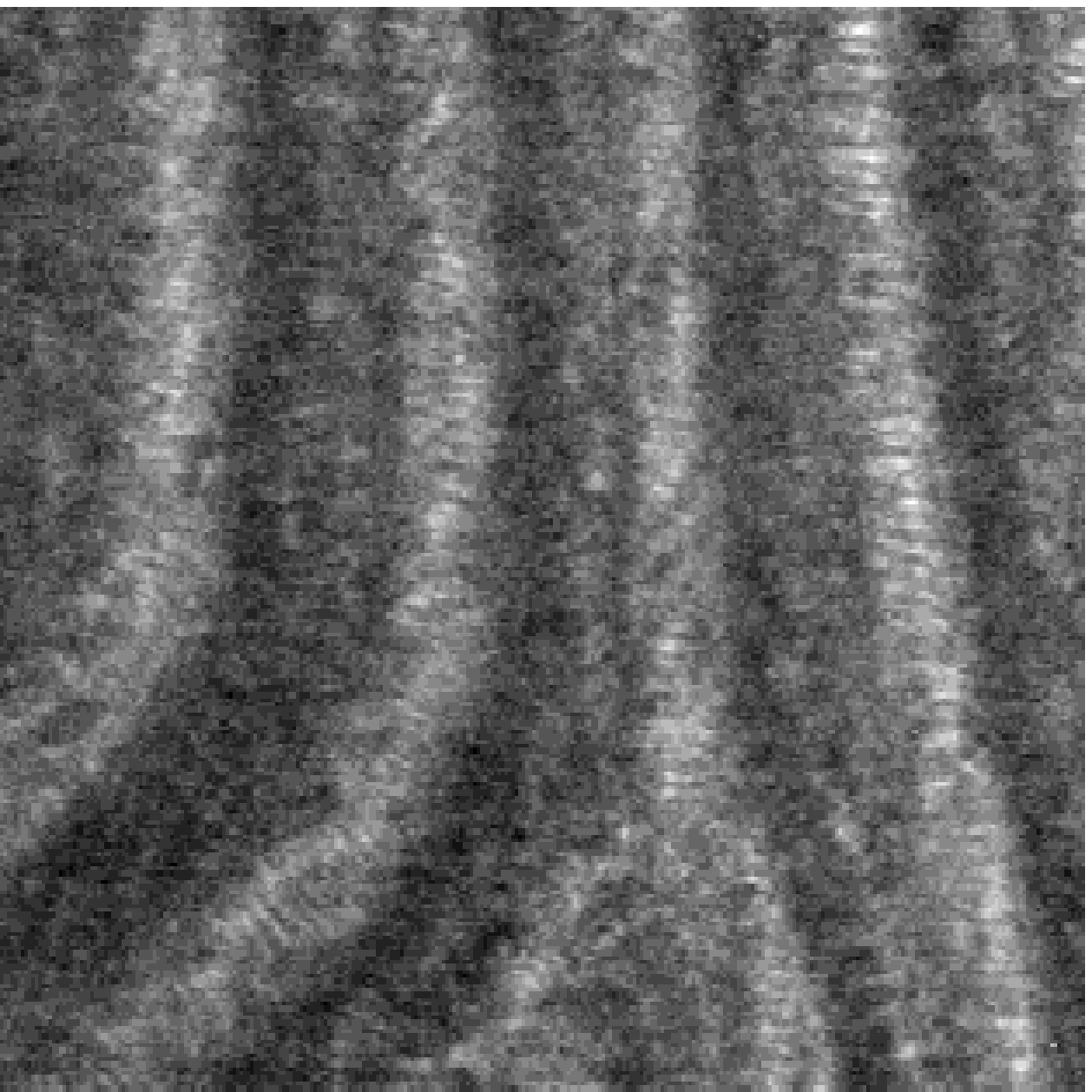,height=10cm}

\vfill
Fig. 6
\pagebreak

c)\psfig{figure=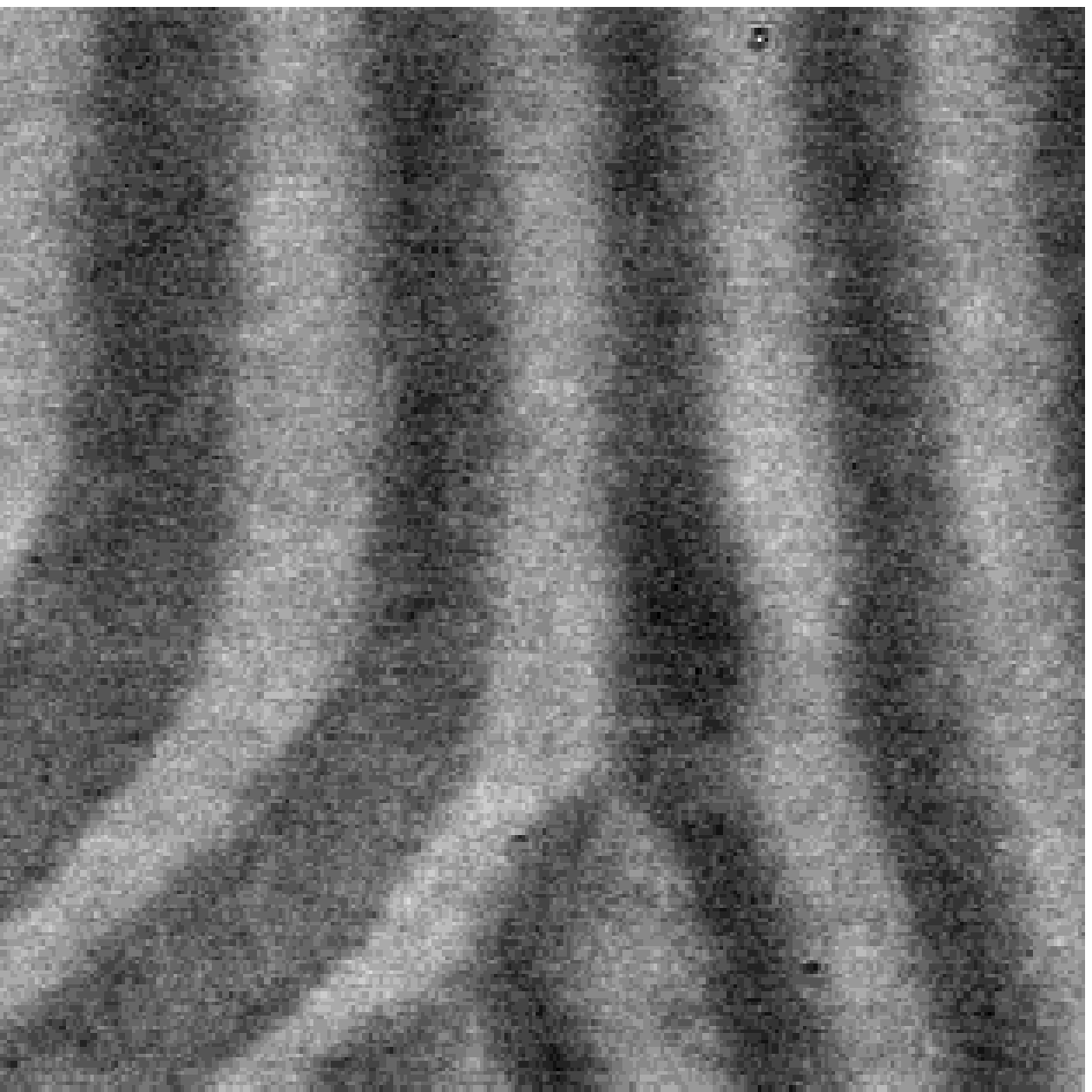,height=10cm}

d)\psfig{figure=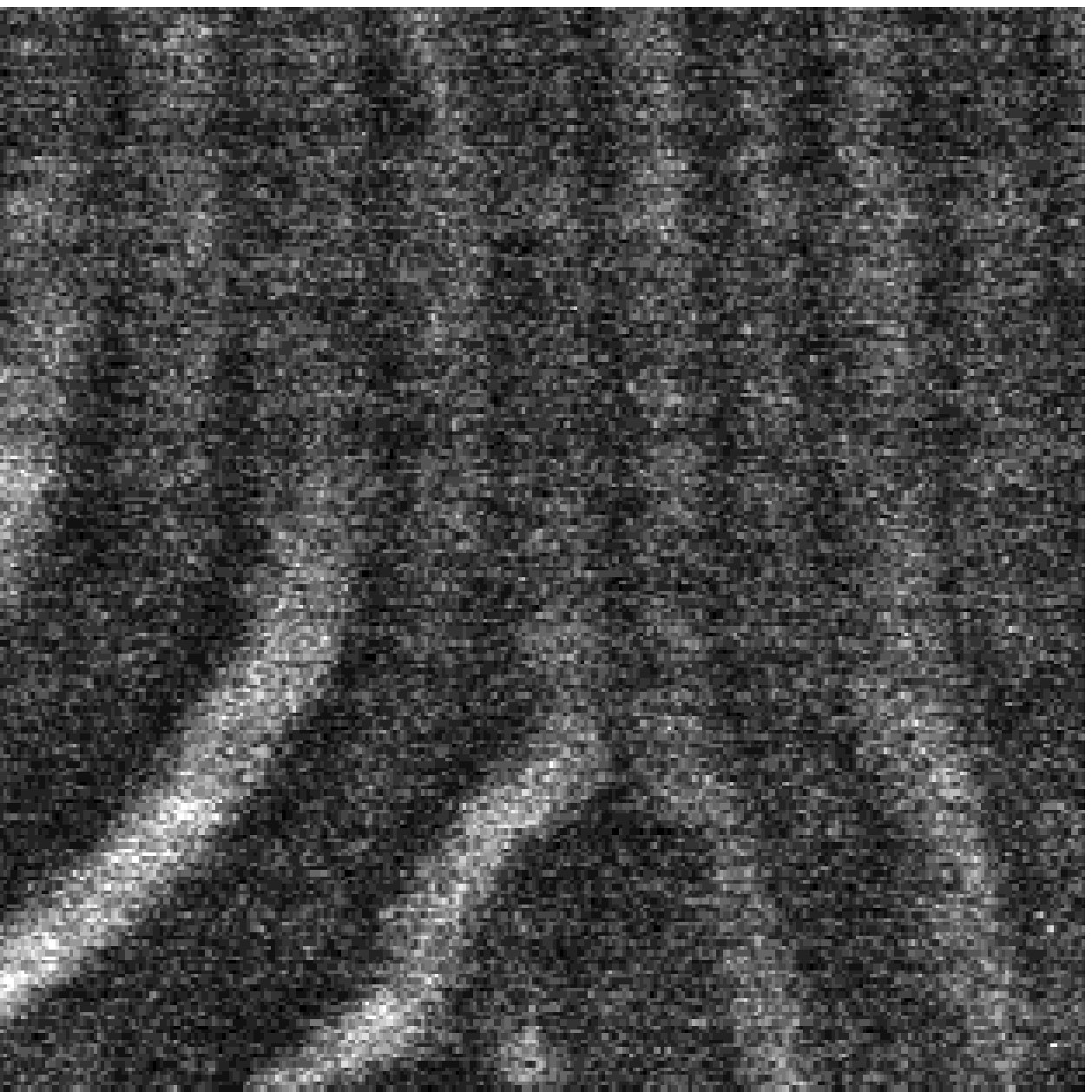,height=10cm}

\vfill
Fig. 7
\pagebreak

\psfig{figure=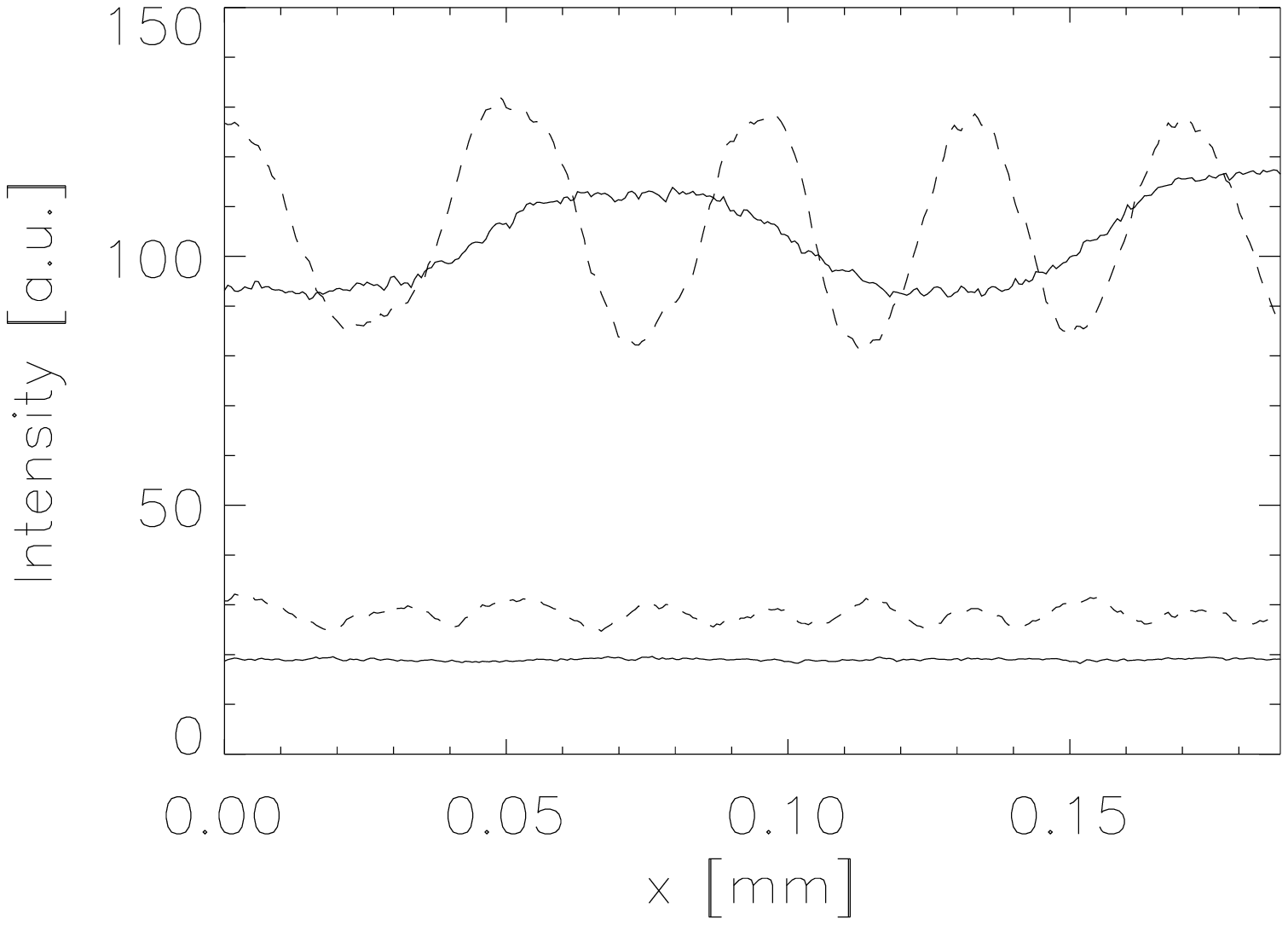}

\vfill
Fig. 8
\pagebreak

\psfig{figure=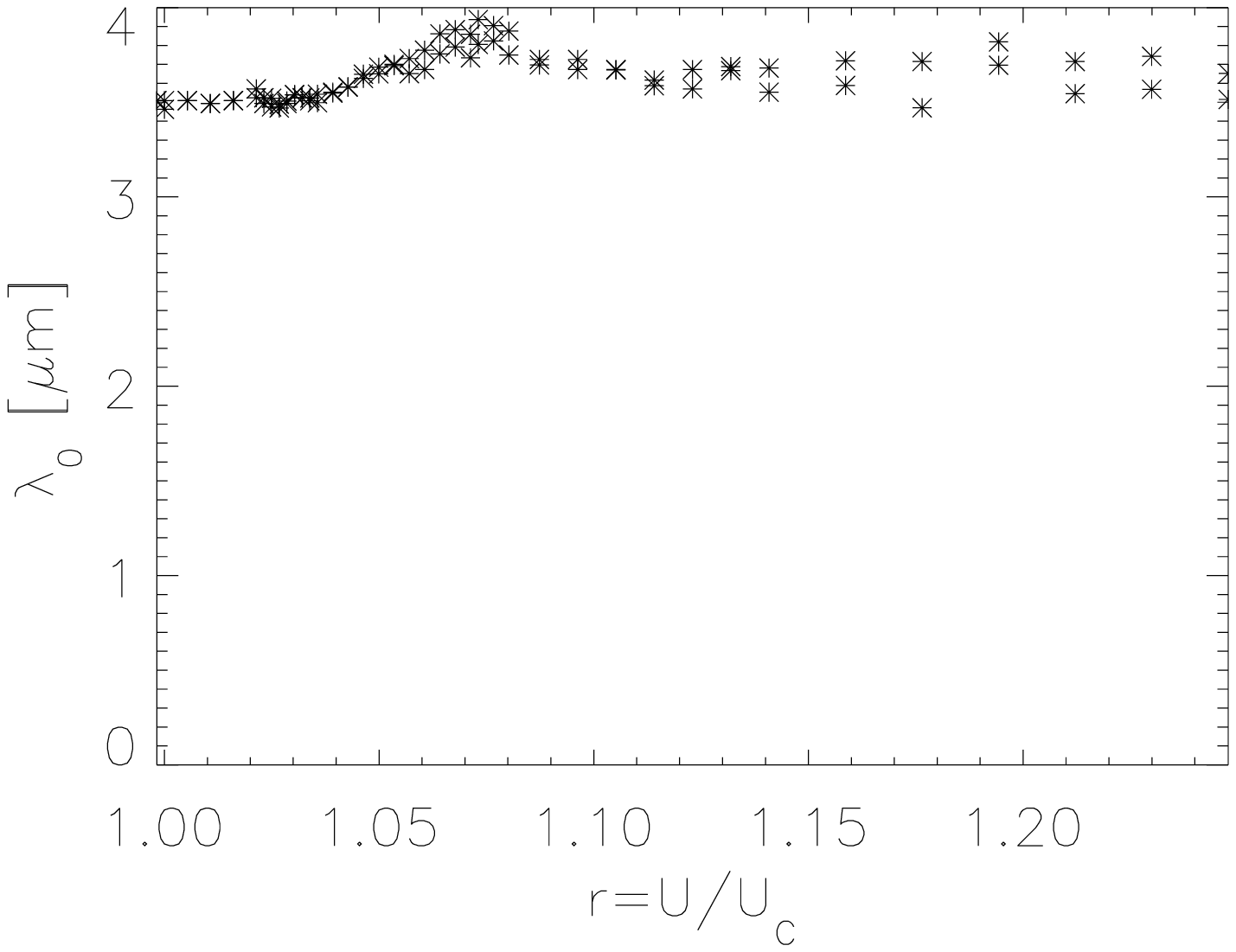}
\vfill
Fig. 9
\pagebreak

\psfig{figure=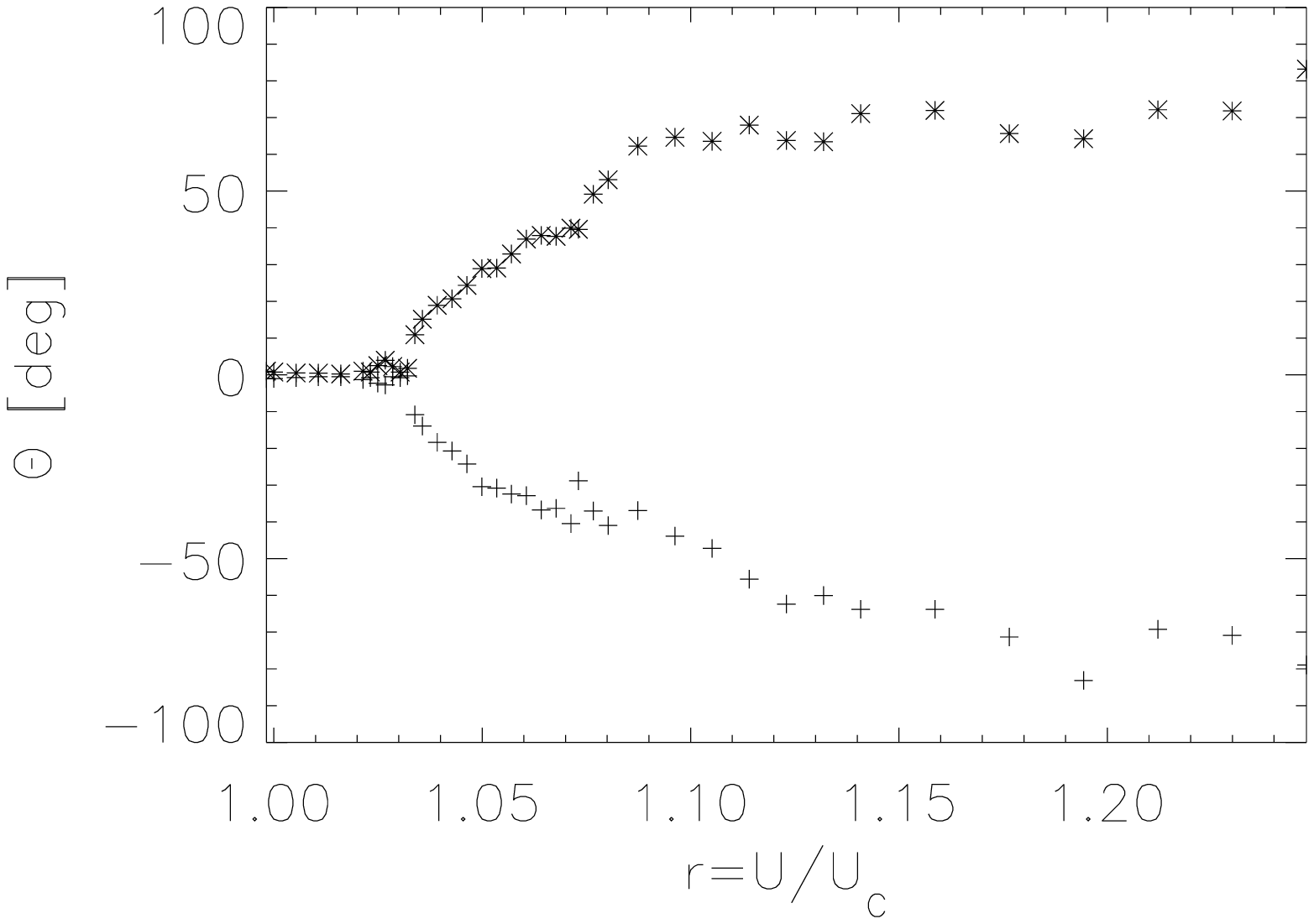}
\vfill
Fig. 10
\pagebreak

\unitlength=1mm
\begin{picture}(150,150)
\put(155,100) {\Large$\phi_m[^\circ]$}
\put(158,24) {\Large 0.0}
\put(155,42) {\Large 18.5}
\put(155,60) {\Large 37.0}
\put(155,78) {\Large 55.5}
\put(-15,0){\psfig{figure=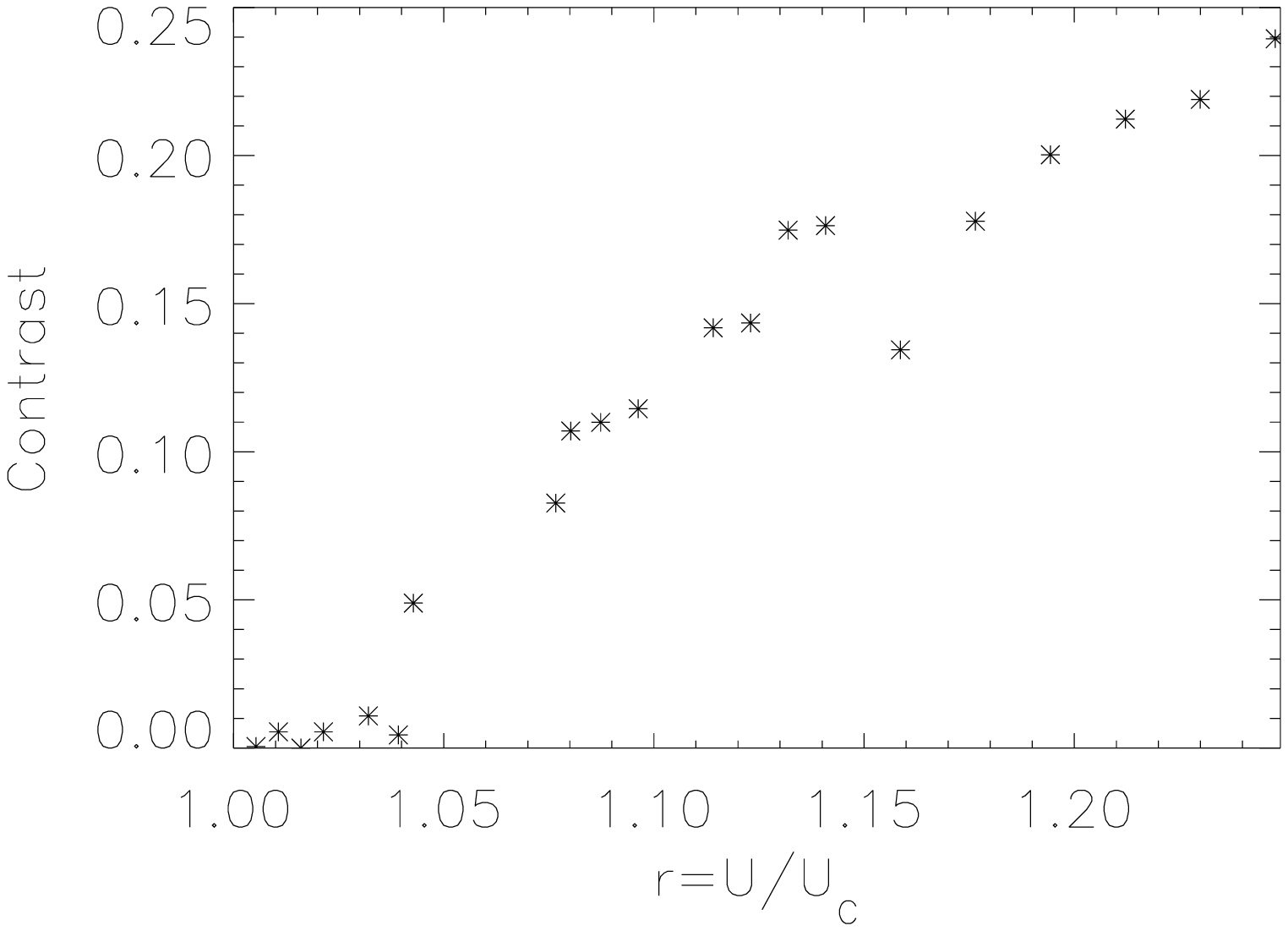}}
\end{picture}
\vfill
Fig. 11
\pagebreak

a)\hspace*{2cm}\psfig{figure=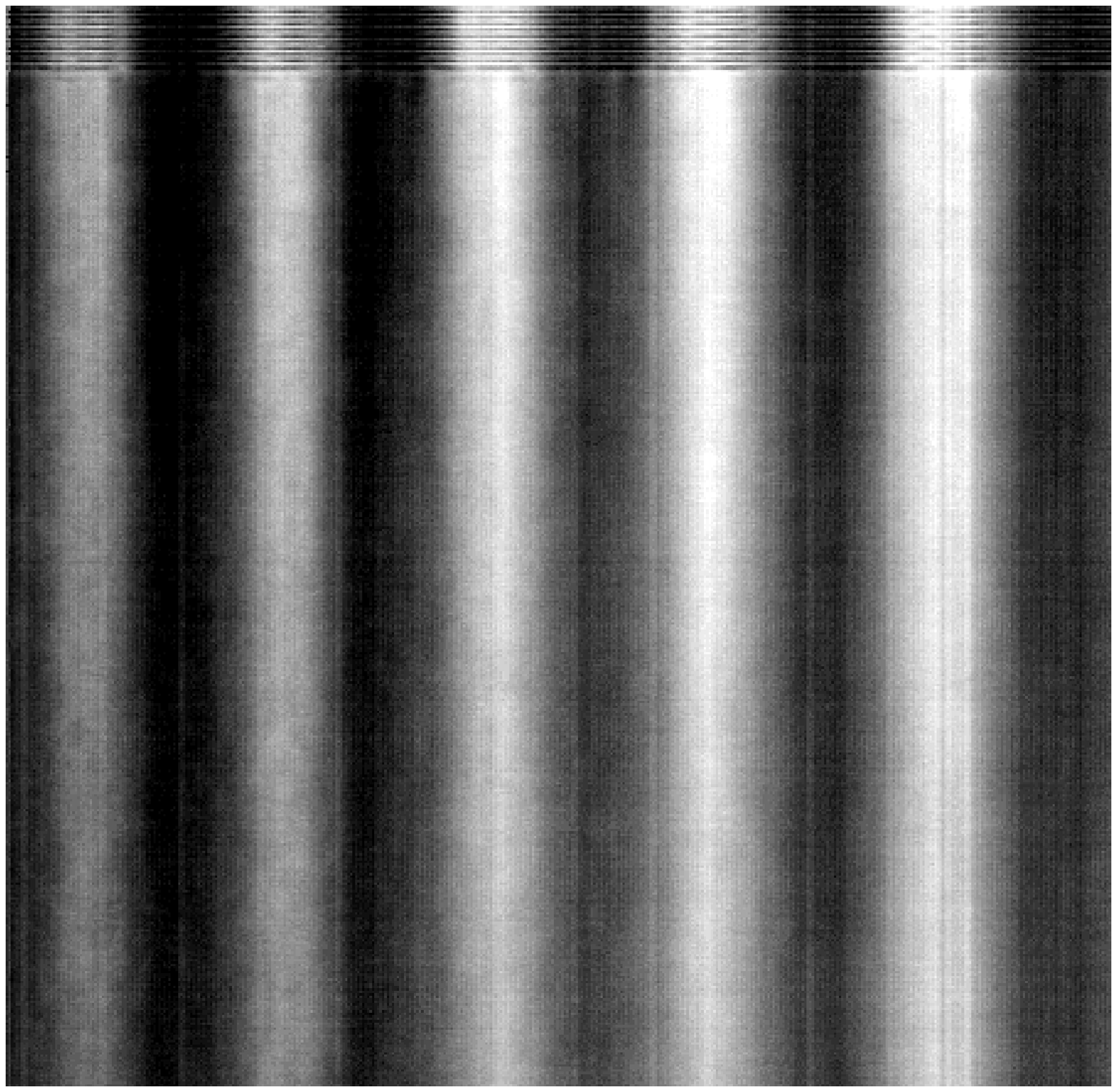,height=10cm}

b)\psfig{figure=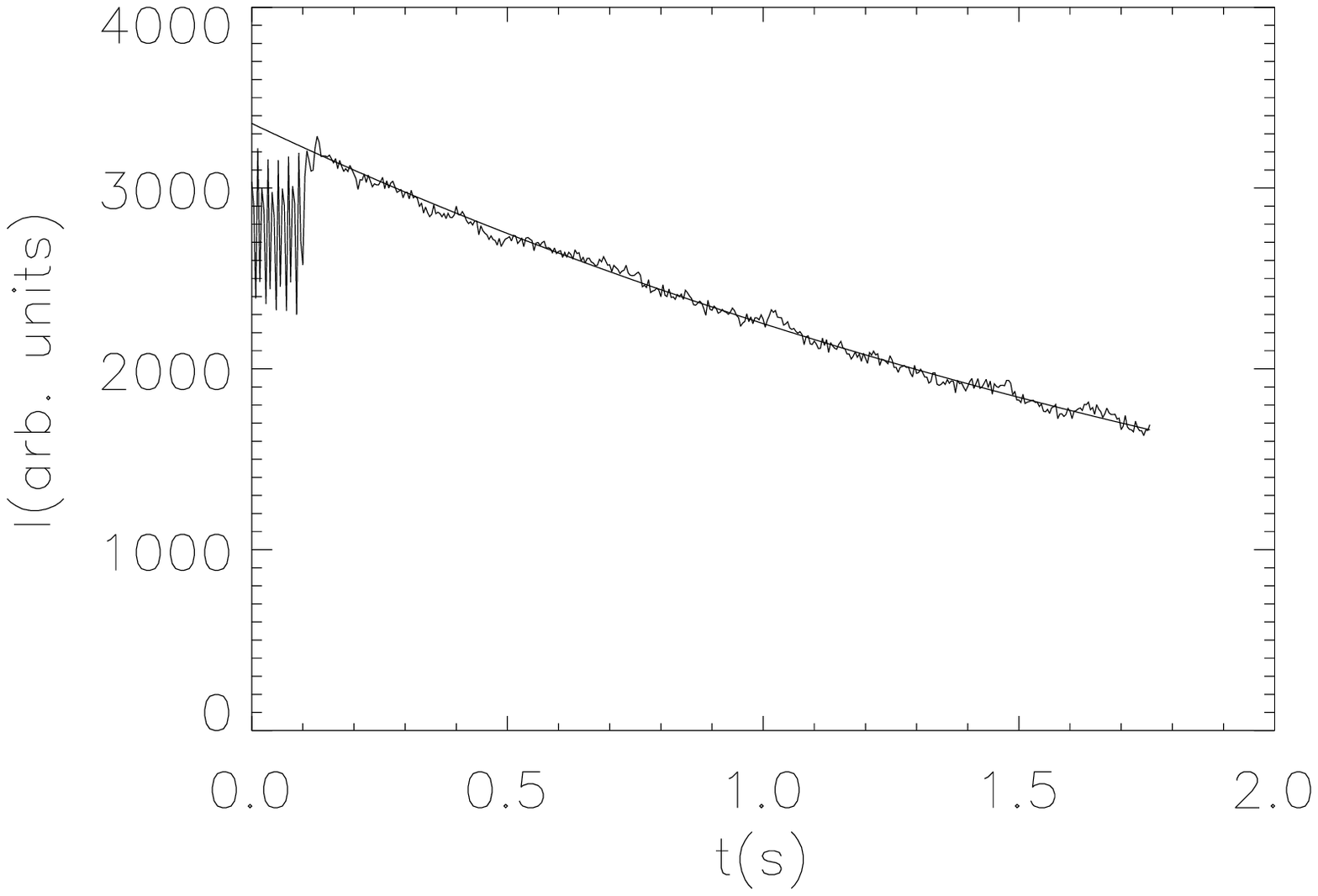,height=10cm}
\vfill
Fig. 12

\end{document}